\documentclass[aps,prl,twocolumn,preprintnumbers,superscriptaddress,floatfix,nofootinbib,notitlepage,showkeys,showpacs]{revtex4-1}

\usepackage[utf8x]{inputenc}
\usepackage{titlesec}
\usepackage{graphicx}
\usepackage{hyperref}
\usepackage{latexsym}
\usepackage{amsmath}
\usepackage{amssymb}
\usepackage{bbm}
\usepackage{ulem}
\usepackage{pdfsync}
\usepackage{epsfig}
\usepackage{epstopdf}
\usepackage{color}
\usepackage{comment}
\usepackage{placeins}
\usepackage{slashed}
\usepackage{mathrsfs}

\newcommand{\be}{\begin{equation}}
\newcommand{\ee}{\end{equation}}
\newcommand{\br}{\begin{eqnarray}}
\newcommand{\bea}{\begin{equation}\begin{aligned}}
\newcommand{\eea}{\end{aligned}\end{equation}}
\newcommand{\er}{\end{eqnarray}}
\newcommand{\ba}{\begin{array}}
\newcommand{\ea}{\end{array}}
\newcommand{\bi}{\begin{itemize}}
\newcommand{\ei}{\end{itemize}}
\newcommand{\bn}{\begin{enumerate}}
\newcommand{\en}{\end{enumerate}}
\newcommand{\bc}{\begin{center}}
\newcommand{\ec}{\end{center}}
\newcommand{\Eq}[1]{Eq.~(\ref{#1})}

\newcommand{\gsim}{\lower.7ex\hbox{$\;\stackrel{\textstyle>}{\sim}\;$}}
\newcommand{\lsim}{\lower.7ex\hbox{$\;\stackrel{\textstyle<}{\sim}\;$}}

\begin{document}
\rightline{CERN-TH-2018-035}

\title{Light primordial exotic compact objects as all dark matter}

\author{Martti Raidal}
\email{martti.raidal@cern.ch}
\affiliation{Theoretical Physics Department, CERN, CH-1211 Geneva 23, Switzerland}
\affiliation{NICPB, R\"avala 10, 10143 Tallinn, Estonia}
\author{Sergey Solodukhin}
\email{sergey.solodukhin@lmpt.univ-tours.fr}
\affiliation{Theoretical Physics Department, CERN, CH-1211 Geneva 23, Switzerland}
\affiliation{Institut Denis Poisson, UMR CNRS 7013, Universit\'e de Tours, Universit\'e d'Orl\'eans, 
Parc de Grandmont, 37200 Tours, France}
\author{Ville Vaskonen}
\email{ville.vaskonen@kbfi.ee}
\affiliation{NICPB, R\"avala 10, 10143 Tallinn, Estonia}
\author{Hardi Veerm\"ae}
\email{hardi.veermae@cern.ch}
\affiliation{Theoretical Physics Department, CERN, CH-1211 Geneva 23, Switzerland}
\affiliation{NICPB, R\"avala 10, 10143 Tallinn, Estonia}

\begin{abstract}
The radiation emitted by horizonless exotic compact objects (ECOs), such as wormholes, 2-2-holes, fuzzballs, gravastars, boson stars, collapsed polymers, superspinars etc., is expected to be strongly suppressed when compared to the radiation of black holes. If large primordial curvature fluctuations collapse into such objects instead of black holes, they do not evaporate or evaporate much slower than black holes and could thus constitute all of the dark matter with masses below $M < 10^{-16}M_\odot.$  We reevaluate the relevant experimental constraints for light ECOs in this mass range and show that very large new parameter space down to ECO masses $M\sim 10$~TeV opens up for light primordial dark matter. A new dedicated experimental program is needed to test this mass range of primordial dark matter.
\end{abstract}

\maketitle


\section{Introduction}
\label{sec:Introduction}

According to the original idea by Hawking~\cite{Hawking:1971ei}, large primordial fluctuation could collapse into primordial black holes (PBHs) when entering into the horizon during radiation dominated era. Consequently, the Universe could be filled with light PBHs with mass $M \gsim 10^{-5}$g (corresponding to the Schwarzschild radius of one Planck length) which could constitute the cosmological dark matter (DM)~\cite{Carr:1974nx,1975ApJ...201....1C,1975A&A....38....5M,CHAPLINE:1975aa}. However, the proposal of Hawking radiation~\cite{Hawking:1974rv,Hawking:1974sw} changed this reasoning dramatically. The light PBHs should evaporate and inject extra photons into the Universe that have not been observed~\cite{Carr:2009jm,Poulin:2016anj}. Combining all existing experimental constraints~\cite{Carr:2009jm}, no PBHs with masses smaller than $M\lsim 10^{-16}M_\odot\sim 10^{17}\rm{g}$ should exist today in any relevant cosmological abundance. Above this limit, in the mass range $10^{-16} \lsim M/M_\odot \lsim 10^{5}$, the lensing limits~\cite{Barnacka:2012bm,Niikura:2017zjd,Tisserand:2006zx,Allsman:2000kg}, various astrophysical and cosmic microwave background constraints~\cite{Capela:2013yf,Graham:2015apa,Koushiappas:2017chw,Brandt:2016aco,Monroy-Rodriguez:2014ula,Ali-Haimoud:2016mbv} as well as the PBH merger rate estimates~\cite{Raidal:2017mfl,Ali-Haimoud:2017rtz} imply that the PBHs cannot be the dominant DM component~\cite{Carr:2017jsz}.

Those considerations are based on predictions of the general relativity (GR). Theories beyond GR that attempt ultraviolet completion of gravity contain new solutions for exotic compact objects (ECOs), such as wormholes~\cite{Damour:2007ap, Berthiere:2017tms,Hohmann:2018shl}, 2-2-holes~\cite{Holdom:2016nek}, fuzzballs~\cite{Mathur:2005zp,Mathur:2008nj}, gravastars~\cite{Mazur:2004fk,Mottola:2006ew}, boson stars~\cite{Kaup:1968zz,Ruffini:1969qy}, black stars~\cite{Barcelo:2009tpa,Barcelo:2007yk}, superspinars~\cite{Gimon:2007ur}, collapsed polymers~\cite{Brustein:2016msz} etc. (see Ref.~\cite{Cardoso:2017cqb} for the complete list of known proposals), whose properties mimic those of  black holes if only long-distance gravitational effects are considered. Studying how to distinguish ECOs from black holes is currently one of the most active research fields~\cite{Berti:2006qt,Macedo:2013qea,Konoplya:2016hmd,Nandi:2016uzg,Chirenti:2016hzd,Barcelo:2017lnx,Brustein:2017koc}. The new physics signatures that allow one to discriminate ECOs from black holes in the binary coalescence, such as the events observed by LIGO~\cite{Abbott:2016blz,Abbott:2016nmj},  include late in-spiral tidal effects and post-merge ring-down tests. In the case of the latter,  the key point is that the absence of horizon of ECOs generates new effects~\cite{Cardoso:2016rao,Cardoso:2016oxy}. For example, if these coalescing objects are ECOs instead of black holes, gravitational wave echoes following the ring-down phase should be present~\cite{Cardoso:2017cqb}.

As there is no observational evidence for Hawking radiation, the experimental status of Hawking's formula for the temperature remains unclear. Therefore, it is well motivated to study how its modifications will affect the cosmological constraints. In this regard, the aim of this work is to study the constraints arising from the radiation of ECOs. Although in the absence of a horizon the usual Hawking mechanism does not apply, the ECOs may still radiate~\cite{Damour:2007ap, Berthiere:2017tms}. To the best of our knowledge, the radiation of ECOs has been studied only in the case of the Damour-Solodukhin wormhole for which the luminosity was found to differ drastically from black holes~\cite{Berthiere:2017tms}.  To cover a wider range of possible modifications we consider exponential-law and power-law changes to the Hawking temperature. We find that such modifications can open a new, wide mass window in which all the DM consists of light primordial ECOs. Indeed, if large primordial curvature fluctuations collapse directly into primordial ECOs, which in theories beyond GR can be as fundamental as black holes, those light primordial objects do not evaporate during the Hubble time and should be present today. Thus, the light primordial ECOs are perfect candidates for the cold collisionless DM of the Universe.  

To achieve this goal we first present a model-independent phenomenological parametrization of the effectiveness of the radiation in the case of ECOs.  After that, we revise the constraints of Ref.~\cite{Carr:2009jm} and show that the astrophysical bounds on light primordial ECOs are lifted. This opens up an entirely new mass window for the primordial DM. Dedicated observations and experiments are needed to test the new mass window for primordial ECOs.

\section{General considerations for ECO evaporation rate}
\label{sec:General}

The unknown quantum gravity effects are expected to modify the Hawking radiation of light black holes by a factor of few~\cite{Helfer:2003va}. At the same time, the possible emitted radiation rate for wormholes is exponentially suppressed because of the absence of a horizon~\cite{Damour:2007ap, Berthiere:2017tms}. Notice that Hawking radiation has never been measured. In order to describe the radiation of as wide a range of ECO candidates as possible, we first present completely model-independent parametrization of the modified radiation effects.

Assuming thermal radiation, the mass dissipation of  a spherical object of radius $r$  is given by
\be \label{dMdt}
\frac{dM}{dt}\simeq-456T^4r^2\, .
\ee
We assume that the radius  $r$ of an ECO is related to its mass $M$ in the same way as in GR,\footnote{This is a very good approximation for most of the ECO candidates that mimic black holes~\cite{Cardoso:2017cqb}.} $r=2M/M^2_{\rm P}$, where $M_{\rm P}$ is the Planck mass, and only their temperature is different from the Hawking temperature by a mass-dependent factor $F(M/\Lambda)$ as
\be \label{temp}
T=\frac{M^2_{\rm P}}{8\pi M}  F(M/\Lambda)^{-\frac{1}{4}}.
\ee
Here $\Lambda$ is the characteristic energy scale of the modified theory of gravity beyond GR which can vary from 10~TeV up to the Planck scale, and the function $F(M/\Lambda)$ is to be specified later. The usual black holes correspond to $F(M/\Lambda)=1$, and $F(M/\Lambda)^{-1}=0$ in the case that the ECOs do not evaporate at all ($T=0$). In our numerical examples we shall consider two limiting cases, $\Lambda=M_{\rm P}$ and $\Lambda=10$~TeV. The latter case corresponds to a situation when the fundamental gravity scale is as low as allowed by the current experimental bounds.

Integration of \Eq{dMdt} gives the evaporation time of an object with mass $M$ to be
\bea \label{evaptime}
t(M) &= \int_0^M dm \frac{128\pi^4}{57}\frac{m^2}{M^4_{\rm P}} F(m/\Lambda) \\ 
&\leq \frac{128\pi^4}{171} \frac{M^3}{M^4_{\rm P}} F(M/\Lambda) \simeq 10^{65} \frac{M^3}{M^3_{\odot}}  F(M/\Lambda) \, {\rm yr} .
\eea
We shall use this last inequality in our numerical estimates. The lower bound on the ECO mass comes from the requirement that the evaporation time should be longer that the age of the Universe, $t(M) < 10^{10}$~yr, which implies
\be \label{bound}
\frac{M^3}{M^3_\odot} F(M/\Lambda)>10^{-55}.
\ee

Before considering some possible forms for the function $F$, we would like to stress at this point that the assumed temperature dependence on function $F$ in \Eq{temp} is motivated by the known behavior of purely gravitational ECOs. For other types of ECOs, such as all the ``stars" listed in the Introduction, the possible forms of  function $F$ may vary depending on the associated physics. In this paper we are mostly interested in purely gravitational ECOs whose production mechanisms are similar to PBH production and which may, therefore, mimic the PBH DM. Below we consider the exponential and power-law behavior of $F$. The first one is motivated by the wormhole solution~\cite{Berthiere:2017tms}, and the second one by modifications to black hole evaporation due to unknown quantum gravity effects.

\medskip

\noindent{\bf Exponential law.} For ECOs without a horizon, the natural expectation is that the emitted radiation rate is exponentially suppressed compared to the Hawking radiation of black holes, $F(M/\Lambda) = e^{(M/\Lambda)^n-1}$. Taking the logarithm of \Eq{bound}, and dropping the logarithmic term $\ln (M/M_{\rm P})$,
we arrive at a bound
\be \label{bound3}
M>\Lambda \ 137^{1/n}.
\ee

i) The case $\Lambda=M_{\rm P}$ and $n=2$ corresponds to the Damour-Solodukhin wormhole~\cite{Berthiere:2017tms}. In this case the lower bound is of order of few dozen Planck masses,
\be
M> 68 M_{\rm P}\simeq 10^{-38}M_\odot .
\ee

ii) In a general case, unless $n$ is extremely small, the numerical factor in \Eq{bound3} is between 1 and $10^2$, and the ECO mass bound is basically set by the value of $\Lambda$.  For $\Lambda=10$~TeV we find
\be
M>10^{-53}M_\odot ,
\ee
consistent with our estimates after \Eq{bound2}.

\medskip

\noindent{\bf Power law.} To describe ECOs whose evaporation is modified less drastically than for the exponential suppression, a natural choice for $F$ is the power-law dependence, $F(M/\Lambda) = (M/\Lambda)^\alpha$. If $\alpha<-3$ the ECO never completely evaporates (the integral in \Eq{evaptime} does not converge). For $\alpha>-3$ \Eq{bound} implies
\be
M>10^{-\beta(\alpha)}\left(\frac{\Lambda}{M_{\rm P}}\right)^\frac{\alpha}{3+\alpha}M_{\odot}\,, \quad \beta(\alpha)=\frac{55+38\alpha}{3+\alpha}.
\ee
Considering the two limiting cases for the fundamental scale of gravity we obtain:

i) If $\Lambda=M_{\rm P}$ then
\be
M>10^{-\beta(\alpha)}M_\odot .
\ee
The case of usual Hawking temperature corresponds to $\alpha=0$ with the usual bound $M>10^{-18.3}M_\odot$, for $\alpha$ of order unity one has $\beta\approx 23$, and in the limit $\alpha\to\infty$ the bound becomes much weaker, $M>10^{-38}M_\odot \simeq M_{\rm P}$. ECOs evaporate completely faster than the usual black holes if $-3<\alpha<0$, and for example $\alpha=-1$ gives a bound $M>10^{-8}M_\odot$.

ii) If the scale $\Lambda$ is much lower than the Planck scale, the ECO mass bound can be even weaker. For instance, if $\Lambda=10\,{\rm TeV} =10^{-15}M_{\rm P}$, one finds
\be \label{bound2}
M>10^{-\gamma(\alpha)}M_\odot \,,  \quad  \gamma(\alpha)=\frac{55+53\alpha}{3+\alpha}.
\ee
In this case the allowed mass of primordial ECOs can be as low as $M>10^{-53}M_\odot\simeq 10$~TeV, consistent with the cutoff scale $\Lambda$.

\section{Reevaluation of experimental bounds} 
\label{sec:Bounds}

The ECO mass bounds derived in the previous section present rough but robust estimates for the primordial ECO DM mass limits. More rigorously,  experimental constraints on the evaporating ECOs arise from the big bang nucleosynthesis (BBN), distortions of the cosmic microwave background, reionization of the Universe, injection of extra entropy, possible modifications of baryogenesis, generation of large positron and antiproton fractions in the cosmic ray fluxes and, above all, from the galactic and extragalactic $\gamma$-ray background measurements. Among the many observables, in practice, only two of them turned out to be relevant for constraining the light, $M\lsim 10^{-16}M_\odot$, PBH and ECO abundance. For the PBHs and ECOs that have completely evaporated by now, the most stringent bounds arise from {\it secondary} $\gamma$-ray flux from their evaporation, and, at even smaller masses, from BBN~\cite{Carr:2009jm}. These bounds constrain the primordial power spectrum at very small scales~\cite{Josan:2009qn,Carr:2017edp,Cole:2017gle}. However, so light ECOs cannot constitute the present DM abundance, and are not of interest for us. For the light ECO DM that exists today, the most stringent bounds arise from the measurements of the extragalactic $\gamma$-ray background. In fact,  the {\it primary} $\gamma$-ray flux from their evaporation turns out to be the only relevant process to consider~\cite{Carr:2009jm}. The extragalactic $\gamma$-ray flux plays a very important role also in constraining properties of annihilating or decaying weakly interacting massive particles~\cite{Hutsi:2010ai,Finkbeiner:2010sm} (for a review see~\cite{Cirelli:2010xx}). Here we  reevaluate the  extragalactic $\gamma$-ray constraints for generic ECOs with a modified radiation rate. We assume that the ECOs have a single mass. The constraints derived here can be generalized to wider mass distributions e.g. using the methods of Ref.~\cite{Carr:2017jsz}.

The present-day primary photon flux produced by evaporating ECOs is a superposition of the instantaneous emissions from all previous epochs. The emission rate per volume at cosmological time $t$ is given by
\be
\frac{{\rm d}n_\gamma}{{\rm d}t}=n_{\rm ECO}(t) E_\gamma \frac{{\rm d}{\dot N}}{{\rm d}E_\gamma}(E_\gamma,M)\,,
\label{n}
\ee
where $n_{\rm ECO}(t)$ is the ECO number density which determines the fraction of DM in ECOs today at $t=t_0$, $f_{\rm ECO} \equiv M n_{\rm ECO}(t_0)/\rho_{\rm DM}(t_0)$, $E_\gamma$ is the emitted photon energy and ${\rm d}{\dot N}/{\rm d}E_\gamma$ is the rate of photons emitted by an ECO in the energy interval $(E_\gamma, E_\gamma+{\rm d}E)$. The form of this rate is determined by the usual black body radiation to be
\be
\frac{{\rm d}{\dot N}}{{\rm d}E_\gamma}(E_\gamma,M) =  \frac{1}{2\pi} \frac{\Gamma(E_\gamma,M)}{e^{E_\gamma/T}-1}\,,
\label{nN}
\ee
where $\Gamma$ is the absorption coefficient which can be approximated in the high-energy limit, $E_\gamma\gg T$, as~\cite{Page:1976df}
\be
\Gamma(E_\gamma,M)= 27 E_\gamma^2 G^2 M^2\,.
\label{G}
\ee
The observable primary photon flux $I_{\rm ECO}(E_\gamma,M) \equiv n_\gamma(E_\gamma,M)/(4\pi)$ is obtained by the integrating \Eq{n} over the time. It is crucial to notice that the average energy of the emitted photons is determined by the temperature alone, $E_\gamma^{\rm av}=5.7 T$, and the peak energy is within 7\% of this value~\cite{MacGibbon:1990zk}. Therefore, to a good approximation, the rescaling of the temperature of radiation according to \Eq{temp} will also rescale the predicted $\gamma$-ray flux. 

The relevant data on the extragalactic $\gamma$-ray background starts at MeV and extends to several hundreds of GeV. To quantify how the bounds depend on the function $F(M/\Lambda)$ in \Eq{temp} we express the flux in terms of the extragalactic $\gamma$-ray flux for PBHs. Applying the temperature rescaling for ECOs, $T = T_{\rm PBH} F(M/\Lambda)^{-1/4}$, together with $T_{\rm PBH} \propto M^{-1}$ to Eqs.~\eqref{n}-\eqref{G} we obtain the relation
\be \label{IECO}
	I_{\rm ECO}(E_\gamma,M) = \frac{I_{\rm PBH}(E_\gamma\,,M F(M/\Lambda)^\frac14)}{F(M/\Lambda)^{\frac14}}.
\ee

The amplitude of the flux is directly proportional to $f_{\rm ECO}$. Thus, the experimental bounds on the $\gamma$-ray flux, $I(E_\gamma,M) < I_{\rm max}(E_\gamma)$, constrain the fraction of DM in ECOs, $f_{\rm ECO}(M) < \min_{E_\gamma} \left[I_{\rm max}(E_\gamma)/I(E_\gamma,M)\right] \equiv f_{\rm ECO}^{\rm max}(M)$. The minimum is attained at the peak energy which is proportional to the temperature. By using \Eq{IECO} we can then relate the maximal allowed fraction of DM in ECOs at given mass $M$ to the maximal PBH DM fraction,
\be
f_{\rm ECO}^{\rm max}(M) = F(M/\Lambda)^\frac14 f_{\rm PBH}^{\rm max}\left(F(M/\Lambda)^\frac14 M\right) \,.
\ee
We use $f_{\rm PBH}^{\rm max} = 3.5\times 10^{55} (M_{\rm PBH}/M_\odot)^{3.4}$ up to arbitrary small masses, though, in particular, above $E_\gamma\sim 100\,{\rm GeV}$ the constraints on the $\gamma$-ray flux get significantly weaker~\cite{Hill:2018trh}. However, in the region where this happens the $\gamma$-ray constraint is practically negligible, as we will see in the next section.

\section{Results} 
\label{sec:Results}

\begin{figure}[t]
\begin{center}
\vspace{-1.2mm}
\includegraphics[width=.432\textwidth]{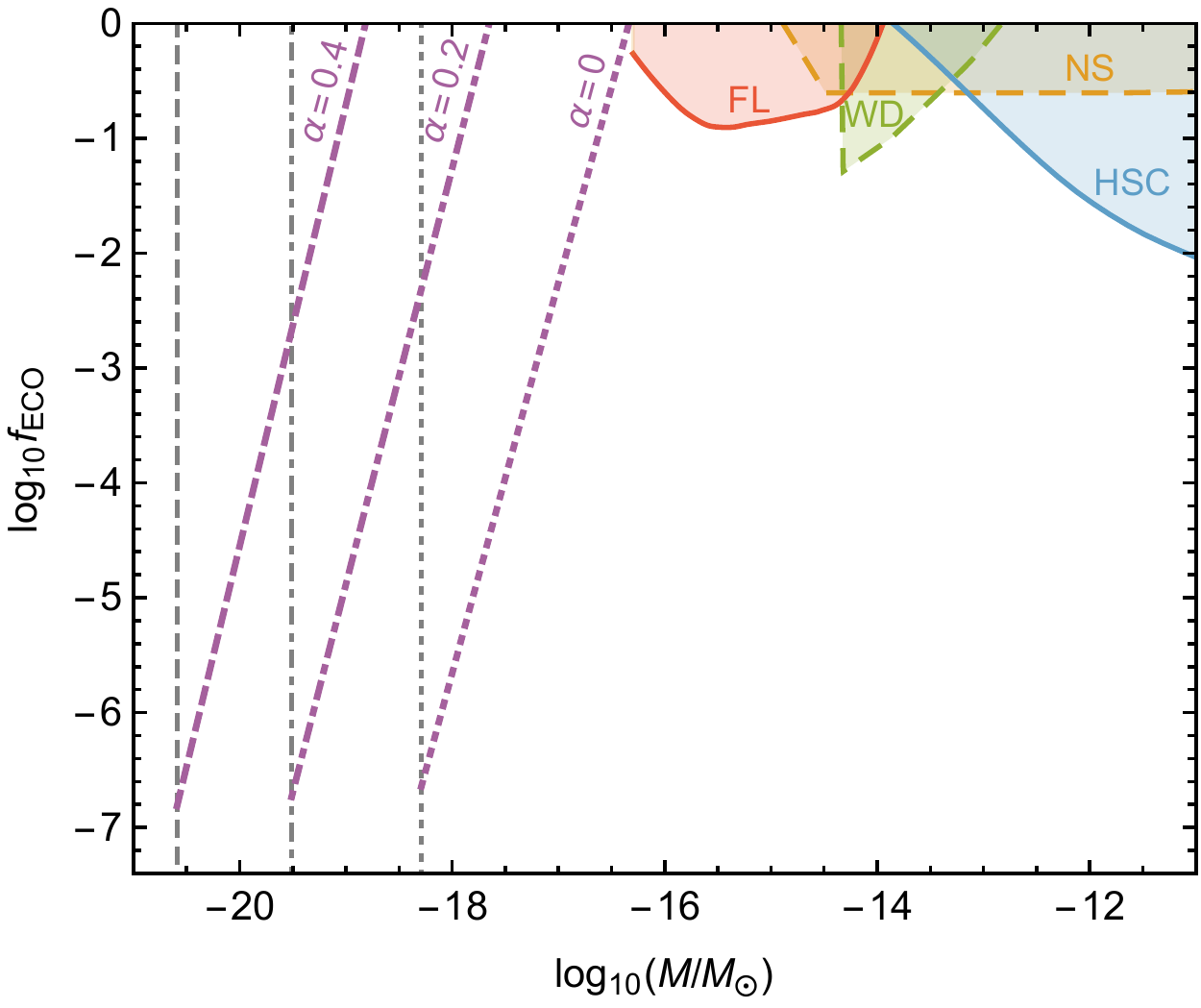}
\caption{Constraints on a fraction of DM in ECOs, $f_{\rm ECO}$, as a function of ECO mass $M$. The dotted, dotted-dashed, and dashed purple lines show upper bounds on $f_{\rm ECO}$ from extragalactic $\gamma$-ray measurements for $F(M/\Lambda)=(M/\Lambda)^\alpha$ with $\Lambda=M_{\rm P}$ and values of $\alpha$ as  indicated in the figure. The vertical gray lines show the corresponding bounds arising from the ECO lifetime alone. The colored regions show constraints from femtolensing (FL)~\cite{Barnacka:2012bm}, white dwarfs (WD)~\cite{Graham:2015apa}, neutron stars (NS)~\cite{Capela:2013yf} and microlensing (HSC)~\cite{Niikura:2017zjd}. For ECOs whose radiation is exponentially suppressed, the mass bounds are lowered by another 30 orders of magnitude.}
\label{fig:constr}
\end{center}
\end{figure}

Using the results above, we plot in Fig.~\ref{fig:constr} the upper bounds on the fraction of DM in ECOs, $f_{\rm ECO}$, arising from the extragalactic $\gamma$-ray measurements as functions of ECO mass $M$ assuming the power-law function $F(M/\Lambda) = (M/\Lambda)^\alpha$, where the values of $\alpha$ are presented in the figure. For the values of $f_{\rm ECO}\approx 1$ in which we are interested in this paper, the bounds from extragalactic $\gamma$-ray measurements are up to two orders of magnitude more stringent than the ones derived from the ECO lifetime, also presented in Fig.~\ref{fig:constr}. The main result is that already for small values of $\alpha>0$, a new parameter space opens up where there is no experimental constraints for the ECO DM abundance.  Therefore, all of the DM can be in the form of objects that radiate less effectively than the classical PBHs, either in ECOs  or, for small values of  $\alpha$, in PBHs where Hawking radiation is modified by unknown quantum gravity effects.

Fig.~\ref{fig:constr} illustrates bounds on ECO DM for small deviations from the Hawking radiation. If the emitted radiation is exponentially suppressed, as is expected for the horizonless ECOs such as wormholes, the mass bounds are further lowered by 30 orders of magnitude.

\begin{figure}[t]
\begin{center}
\includegraphics[width=.44\textwidth]{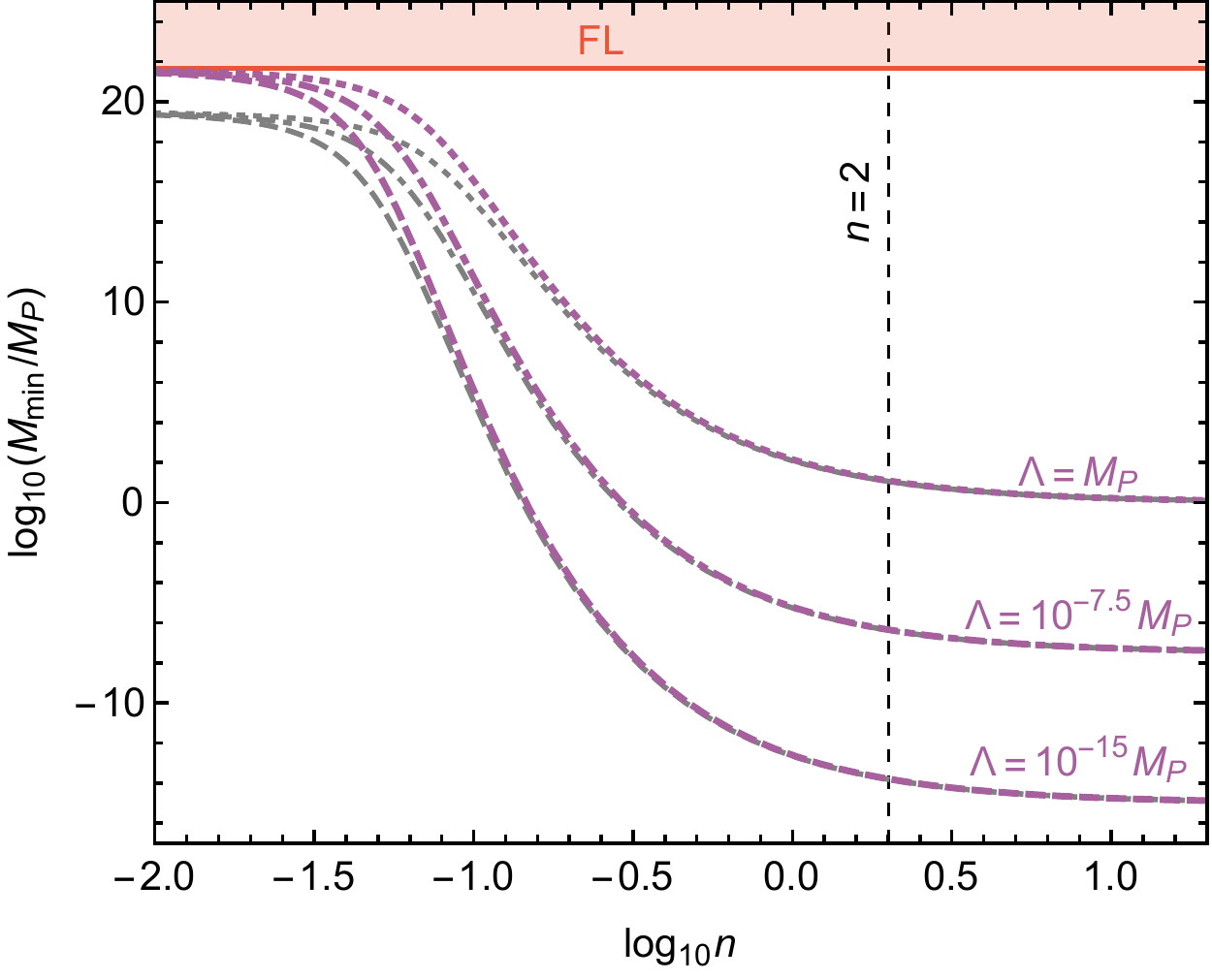}
\caption{The purple lines show the smallest allowed ECO DM mass, $M_{\rm min}$, for $f_{\rm ECO}=1$ and $F(M/\Lambda)=e^{(M/\Lambda)^n-1}$ as a function of $n$ for three different values of $\Lambda$. The gray lines depict the corresponding bounds from the ECO lifetime, and the red region is excluded by the femtolensing (FL) constraint~\cite{Barnacka:2012bm}. The vertical line highlights the case $n=2$ corresponding to the wormhole solution~\cite{Berthiere:2017tms}.}
\label{fig:mmin}
\end{center}
\end{figure}

To study which range of ECO masses are reachable by our considerations, we plot in Fig.~\ref{fig:mmin} lower bounds on the smallest allowed ECO mass $M_{\rm min}$ for which all DM can be in ECOs, $f_{\rm ECO}=1$,  as functions of $n$ for the exponentially suppressed ECO radiation rate $F(M/\Lambda)=e^{(M/\Lambda)^n-1}$. This case is motivated by the wormhole solution~\cite{Berthiere:2017tms} that is highlighted in the figure. We see that the results depend strongly on the cutoff scale $\Lambda.$ Because the scale of the UV theory of gravity is unknown, stable ECO masses as low as 10~TeV are possible. Therefore, the stable ECOs that constitute 100\% of the DM of the Universe need not be macroscopic objects like PBHs. Instead, they can be more exotic solutions with masses just above the present reach of the LHC and other particle physics experiments.

\section{Discussion and Conclusions} 
\label{sec:Conclusions}

We demonstrated that light primordial ECOs or PBHs with modified Hawking radiation at masses below  $10^{-16}M_\odot$ can constitute 100\% of the DM of the Universe without contradicting any experimental bound. The key point that allowed us to reach this result is the absence of a horizon for ECOs, which drastically modifies the emitted radiation rate of those objects compared to PBHs. As a result, the ECO lifetime is significantly prolonged allowing for the existence of very light primordial objects.  We showed that, similarly to the PBHs, the most stringent lower bounds on primordial ECO masses arise from the extragalactic $\gamma$-ray measurements. As is evident from Figs.~\ref{fig:constr} and \ref{fig:mmin}, a new, very large mass window is opened for the primordial ECO DM that mimics the PBH DM. In the extreme cases, when the fundamental cutoff scale of gravity is much below the Planck scale, the primordial ECOs can be as light as 10~TeV, resembling particles rather than macroscopic objects. The most important conclusion, therefore, is that new dedicated observations and experiments are needed to test this mass region of primordial DM.

\medskip

\noindent{\bf Acknowledgements} M.R. thanks the organizers of  COST action CA16104 meeting in Malta where part of the ideas presented in this paper were developed. This work was supported by the grants IUT23-6 and by EU through the ERDF CoE program grant TK133 and by the Estonian Research Council via the Mobilitas Plus grant MOBTT5.

\vspace{-5mm}
\bibliography{citations}

\end{document}